\documentclass[12pt]{article}

\newcommand{\eq}{\begin{equation}}
\newcommand{\eqx}{\end{equation}}
\newcommand{\eqn}{\begin{eqnarray}}
\newcommand{\eqnx}{\end{eqnarray}}
\newcommand{\f}[2]{\frac{#1}{#2}}

\newcommand{\tr}{\mbox{\rm tr}\,}

\newcommand{\om}{\omega}
\newcommand{\dl}{\delta}
\renewcommand{\th}{\theta}

\newcommand{\al}{\alpha}

\newcommand{\kap}{\kappa}

\newcommand{\cor}[1]{\left\langle{#1}\right\rangle}

\usepackage{graphics}
\usepackage{psfig}
\usepackage{amssymb}

\newcommand{\rf}[1]{(\ref{#1})}
\newcommand{\beq}{\begin{equation}}
\newcommand{\eeq}{\end{equation}}
\newcommand{\bea}{\begin{eqnarray}}
\newcommand{\eea}{\end{eqnarray}}

\newcommand{\m}{\mu}

\renewcommand{\th}{\theta}
\newcommand{\oh}{\frac{1}{2}}

\newcommand{\ra}{\rangle}
\newcommand{\la}{\langle}
\newcommand{\prt}{\partial}

\newcommand{\equ}{\!=\!}
\newcommand{\pl}{\!+\!}

\begin{document}

\begin{center}
\vspace{24pt}
{ \large \bf Quantum rolling tachyon}

\vspace{30pt}

{\sl Jan Ambj\o rn}$\,^{a,c}$ and {\sl Romuald A. Janik}$\,^{b}$,

\vspace{24pt}
{\footnotesize

$^a$~The Niels Bohr Institute, Copenhagen University\\
Blegdamsvej 17, DK-2100 Copenhagen \O , Denmark.\\
{ email: ambjorn@nbi.dk}\\

\vspace{10pt}

$^b$~Institute of Physics, Jagellonian University,\\
Reymonta 4, PL 30-059 Krakow, Poland.\\
{ email: janik@th.if.uj.edu.pl}\\

\vspace{10pt}

$^c$~Institute for Theoretical Physics, Utrecht University, \\
Leuvenlaan 4, NL-3584 CE Utrecht, The Netherlands.\\

\vspace{10pt}

}
\vspace{48pt}

\end{center}


\begin{center}
{\bf Abstract}
\end{center}

We consider the quantum treatment of the rolling tachyon background
describing the decay of D-branes in the limit of weak string
coupling. We focus on the propagation of an open string in the
fluctuating background and show how the boundary string action is modified 
by quantum effects. A bilocal term in the boundary action is generated
which, however, does not spoil the vanishing of the $\beta$ function
at one loop. The propagation of an open string for large times is
found to be very strongly suppressed.

\vspace{12pt}
\noindent


\newpage

\subsection*{Introduction}

D-branes are maybe the most fascinating and still somewhat 
mysterious objects in string theory. In particular 
the precise quantum nature of the branes is not entirely 
clear: in perturbative string theory they appear as 
boundary conditions for the open string, while 
in the low energy effective field theories or string field theory
(SFT) they appear 
as solitonic solutions to the equations of motion.
The analogy with solitons in ordinary quantum 
field theory is in this respect slightly misleading,
since solitons there never appear as additional 
degrees of freedom. From the SFT perspective one could argue
similarly, yet there are some glitches in the picture -- attempts to find a
solution of SFT describing {\em two} D25 branes have failed so far
\cite{taylor}. So it is not clear what is the complete set of degrees
of freedom.
Indeed sometimes it can be convenient to 
use the solitons as  alternative degrees of freedom
instead of the fundamental fields. However, from 
the point of view of perturbative string theory, D-branes
seem more like heavy (charged) particles which 
in the end have to be included as full quantum 
fields in a consistent quantum theory.
The question of how in detail to quantize the D-branes
is still open but has to be eventually addressed in order
to have a complete theory.

One meets (mild) versions of the quantum nature of 
the D-branes for instance in the study of D-brane 
recoil, but also in the toy models
encountered in non-critical string theory for $c \le 1$.
In particular the case $c=1$ has been much studied since 
it corresponds to 2d critical string theory in a linear 
dilaton background, and the D0 branes, as noted in \cite{kms,mv},
can be viewed as the eigenvalues 
in the $c=1$ matrix model. With this interpretation the decay of an 
(unstable) D0-brane of the 2d bosonic string theory is described by 
a tachyon moving in a potential (we choose $\al'=1$)
\beq\label{j1}
V = -\f{T^2}{2},
\eeq
and the {\it quantum} description of a single 
D0-brane  is dictated by the Hamiltonian
\beq\label{j2}
H = -\oh \f{d^2}{dT^2}+V(T),~~~~V(T)= -\oh T^2.
\eeq 

The $c=1$ matrix model can  be used to analyze the decay of the 
D0-brane into closed string tachyons. In this calculation 
(as in Sen's original calculation in critical string theory)
one only uses the {\it classical} aspects of the open string theory,
here specifically that a classical solution in the potential \rf{j1}
is given by
\beq\label{j3}
T_{cl}(t)= T_0 \, \cosh t.
\eeq
This is an exact counterpart of the rolling tachyon background
\cite{sen} in critical bosonic strings in 26 dimensions.

In fact, in such a calculation one finds that the number of emitted 
(closed string) particles diverges logarithmically and the expectation
value of the emitted energy diverges linearly. This is true for the 
decay of D0-branes both in the $c=1$ theory and in critical string  
theory. In the $c=1$ theory the situation could be analyzed further
since the quantum theory of the D0-branes is known. Clearly the 
classical picture of D0-brane decay is only approximate, limited by the 
uncertainty principle, and it was suggested in \cite{kms} that 
by considering wave packets rather than the localized eigenvalues
corresponding to \rf{j3} one would avoid the divergence caused 
by the classical open string background. This was verified in detail 
in \cite{aj} and in this article we discuss the extension of the 
results in \cite{aj} to critical bosonic string theory.

\subsection*{The quantum boundary action}

Let us first review the 2d calculation. The amplitude for emission of 
closed string tachyons is given by an expression
\beq\label{j10}
A(E) \sim \lim_{l\to 0} \; ({\rm leg~factor})
\int dt \; e^{i Et} \,\langle \tr e^{-il \Phi (t)}\rangle_M
\eeq
where the so called ``leg factor'' is unimportant to us and 
where  
\beq\label{j11}
\langle \tr e^{-il \Phi (t)}\rangle_M \equiv \int d T \; \rho(T,t)\,
e^{-i l T}, 
\eeq
$\Phi$ is a $N\times N$ matrix (where $N \to \infty$) and  $\rho(T,t)$ 
is the eigenvalue density of $\Phi$.
The contribution from the classical rolling tachyon is obtained by the
substitution $\rho(T,t) \to \delta(T -T_{cl}(t))$. 
Thus one ends up calculating the integral 
\beq\label{j12}
\int dt \;e^{i Et} \,e^{-il T_{cl}(t)}.
\eeq
As shown in \cite{aj} this integral gave  precisely the closed tachyon
emission amplitude \rf{j10} obtained using continuum BCFT methods 
in the case of a decaying classical $D0$-brane.
But in fact the true physical object is the {\it quantum} $D0$-brane,
which is described by a wavefunction $\psi(T,t)$ -- satisfying the
appropriate Schroedinger equation. The closed string emission can then
be calculated using the replacement
\beq\label{j13}
e^{-il T_{cl}(t)} \to \left\langle e^{-il T} \right\rangle_{\psi(t)} \equiv 
\int dT \;|\psi(T,t)|^2 e^{-il T}.
\eeq

Let us now turn to the calculation of the decay of D-branes in
critical (bosonic) string theory. The dominant source of 
D-brane  instability is  the tachyon field mode $T(t,\vec{x})$. It lives 
on the boundary of the world sheet and gives rise to 
a boundary action  
\beq\label{ba}
S_b=\int_B d\tau \; T(t,\vec{x}),
\eeq
where the integration is along the string boundary.
Let us expand the tachyonic field in momentum modes,
where for simplicity we compactify 
the spatial directions to a box with volume
$V=L^{p}$ for a Dp-brane: 
\eq\label{fourier}
T(t,\vec{x}) =\f{1}{(2\pi L)^{\f{p}{2}}} \sum_{\vec{n}}
T^c_{\vec{n}} (t) \cdot \cos\left(2 \pi n \f{x}{L}\right) +
T^s_{\vec{n}} (t) \cdot \sin\left(2 \pi n \f{x}{L}\right)
\eqx
We now need the {\em spacetime} action for the tachyon field. We will
consider just the free quadratic part of Witten's OSFT, thus we assume
that $g_s$ is small. We will then treat this spacetime action
quantum-mechanically. We still expect nontrivial quantum effects even
using just the free SFT. Indeed in \cite{aj}, in the 2d string case,
we observed that quantum effects were very important even before
including the effects of string interactions (the Fermi sea).

Inserting this decomposition into the quadratic string field theory
action one obtains the following:
\eq\label{s_t}
S_T= \sum_{\vec{n}} \int dt \left\{ \f{1}{2}{\dot{T}}^{c,s}_{\vec{n}} (t)^2
-\f{1}{2} (k^2_{\vec{n}}-1) {T^{c,s}_{\vec{n}}} (t)^2 \right\}
\eqx
Thus we get a set of inverted harmonic oscillators (with $k^2<1$) and
a set of normal ones (modes with $k^2>1$). This is just the linearized 
approximation to a genuine string field theoretical description
but will be sufficient for our purpose.

The dominant unstable mode is the spatial constant mode, so we consider 
first that mode. We will return to the nonconstant modes later in the
paper. 
The (now) standard calculation of closed string
emission assumes that we have a {\em classical} tachyonic background
$T_{cl}(t)$, $T_{cl}(t)$ being a solution to the classical equation 
originating from the truncated action \rf{s_t}. Such solutions 
are of the form 
\beq\label{j4}
T_{cl}(t) = T_0\cosh t~~~{\rm or} ~~~~T_{cl}(t)= T_0 e^{t}. 
\eeq


In all calculations so far, whether one uses string field theory and 
the formalism of boundary states \cite{sen} or the conceptually simpler open
string perturbative approach of \cite{finn}, one always treats the 
tachyon profile as given by \rf{j4}, i.e. as being strictly classical.

As mentioned above, one lesson from the 2d critical string 
theory was that one has eventually to promote the classical 
tachyonic background to a {\it quantum state}. One 
(first quantized) way to do that is to  replace the 
classical solution with a {\em quantum} wave function:
\beq\label{j6}
T_{cl}(t) \to \psi(T,t),
\eeq 
where the wave function $\psi(T,t)$ is a solution to the
corresponding quantum Hamiltonian coming from the action
\rf{s_t}, i.e. in the case where we restrict ourselves
to the space-independent tachyon mode precisely \rf{j2}.
This should capture the most important quantum aspects of the D-brane
as long as one does not consider the annihilation and 
creation of D-branes.

Our aim now is to determine how an open string will propagate (and
e.g. eventually emit closed string radiation) in the {\em fluctuating}
quantum background given by $\psi(T,t)$.

An open string moving in the classical open string background
$T_{cl}(t)$ has a boundary action\footnote{Throughout the paper we are
using Minkowski signature.}
\eq
e^{i \int_B d\tau T_{cl}(t(\tau))}
\eqx
Let us recast it in the following form:
\eq\label{j7}
e^{i \int_{t_a}^{t_b} dt\, T_{cl}(t) \left( \f{1}{\dot{t}_1}+\ldots
+\f{1}{\dot{t}_n}\right)} \equiv  e^{i \int_{t_a}^{t_b} dt\, T_{cl}(t) J(t)}
\eqx 
where $t_i$ are the parameterizations of various sectors of the string
boundary and $t_a$ and $t_b$ are the minimum and maximum time
coordinates of the boundary of the string world-sheet and where 
we view $J(t)$ as an external driving force for $T(t)$.
The natural generalization from a classical $T_{cl}(t)$ to a quantum 
state described by the wave function $\psi(T,t)$ is 
\eq\label{qba}
e^{i \int_{t_a}^{t_b} dt\, J(t)T_{cl}(t) } \to 
\cor{e^{i \int_{t_a}^{t_b} dt\, J(t) T(t)}}_\psi,
\eqx
i.e.\ the expectation value of an (inverted) harmonic oscillator 
in the presence of  an external source $J(t)$. This expectation
value can (in principle) be explicitly evaluated since we 
know the propagator $K_J(T_b,t_b;T_a,t_a)$,  
of the (inverted) harmonic oscillator in the presence of an external 
source $J(t)$. For the {\it ordinary} harmonic oscillator with 
cyclic frequency $\om$ one has 
\eq\label{k_j}
K_J(T_b,t_b;T_a,t_a)=\left( \f{\om}{2\pi i \sin \om T} \right)^{\f{1}{2}}
e^{i\left[S^0_{cl}+ \left(T_a J_a +T_bJ_b\right)+J_{ab}\right]} 
\eqx
where $S^0_{cl}(T_a,t_a;T_b,t_b)$ is the classical action
of a harmonic oscillator which at time $t_a$ is at position $T_a$ and 
at time $t_b$ at position $T_b$:
\beq\label{s_cl}
S^0_{cl}(T_a,t_a;T_b,t_b) 
=\f{\om}{2\sin\om T} \left[ (T_a^2+T_b^2) \cos \om T -2
T_a T_b \right],
\eeq 
where $T=t_b-t_a$. Finally $J_a$, $J_b$ and $J_{ab}$ are given by
\eq\label{j_a}
J_a = \int_{t_a}^{t_b} dt\, J(t) 
\f{\sin \om (t_b-t)}{\sin\om T},~~~
J_b =\int_{t_a}^{t_b} dt\, J(t) \f{\sin \om (t-t_a)}{\sin\om T},~~~
\eqx
and
\eq\label{j_ab}
J_{ab} = \f{1}{2} \int dt dt'\, J(t) D_0(t,t') J(t'),
\eeq
where
\eq\label{D0}
D_0(t,t') = \th(t-t') \f{\sin \om(t-t')}{\om} -\f{\sin \om (t-t_a)
\sin \om (t_b-t')}{\om \sin \om T}.
\eqx
$D_0$ is the propagator corresponding to boundary conditions 
$T(t_a)=T(t_b)=0$.
One obtains the corresponding expression 
for the inverted harmonic oscillator by the analytical continuation
$\om \to i \om$. 

We thus have
\eq
\label{e.calc}
\cor{e^{i \int_{t_a}^{t_b} dt\, J(t) T(t)}}_\psi = 
\int dT_a dT_b \,\psi^*(T_b,t_b) K_J(T_b,t_b;T_a,t_a) \psi(T_a,t_a).
\eqx
In the final expression {\em
explicit} dependence on $t_a$ and $t_b$ should cancel and we should
get a functional of $J(t)$. The expression \rf{e.calc} has a natural
interpretation of (the exponential of)
the {\it quantum boundary action} since it describes how the 
exponential of the classical action is changed if $T_{cl}(t) \to 
\psi(T,t)$. 

\subsection*{The boundary action for gaussian wave functions}

In order to exemplify the prescription let us consider a Gaussian wave
packet with the initial condition 
\eq\label{gauss}
\psi(T,0)=\left( \f{a}{\pi} \right)^{\f{1}{4}} e^{-\f{a}{2} (T-T_0)^2}.
\eqx
After a rather long calculation one obtains
\eq\label{e.oscinv}
\cor{e^{i \int_{t_a}^{t_b} dt\, J(t) T(t)}}_\psi = 
e^{iT_0 \int dt\, J(t) \cosh t +\f{i}{2} \int dt dt'
J(t) D_a(t,t') J(t') }
\eqx
or, using the definition of $J(t)$:
\beq\label{qaction}
\cor{e^{i \int_B d\tau  \,T(t(\tau))}}_\psi = 
e^{i \int_B d\tau \, T_0\cosh t(\tau) +\f{i}{2} \int_B\int_B d\tau d\tau'
D_a(t(\tau),t'(\tau')) }.
\eeq
In \rf{e.oscinv} and \rf{qaction} we used the notation
\beq\label{Da}
D_a(t,t')=\th(t-t')\f{\sinh\om (t-t')}{\om} +
i\f{\om^2 \cosh \om t \cosh \om t' + a^2 \sinh \om t \sinh \om t'}{2a\om^2},
\eeq
where we have kept $\om$ (which should be put equal to one) in order
to make clear how this is related to the corresponding expression for 
the ordinary harmonic oscillator by $\om \to i\om$. Note that the
final results for the inverted and normal harmonic oscillators are
{\em not} related by an analytical continuation. This is due to the
wavefunctions appearing in (\ref{e.calc}).
Eq.\ \rf{Da} simplifies if $a\equ \om$: \footnote{For the ordinary 
harmonic oscillator it comes a function  of $t-t'$ only:
\beq\label{Dom1}
D_{\om}(t,t')= \th(t-t')\f{\sin\om (t-t')}{\om} +
i\f{\cos \om (t-t')}{2\om}.
\eeq}
\beq\label{Dom}
D_{\om}(t,t')= \th(t-t')\f{\sinh\om (t-t')}{\om} +
i\f{\cosh \om (t+t')}{2\om}
\eeq

One observes that indeed all explicit reference to $t_a$ and $t_b$ 
has dropped out and that the quantum boundary action corresponding 
to \rf{e.oscinv} or \rf{qaction} 
has the structure of the classical boundary action (the term 
linear in $J(t)$) and a quantum correction (the term quadratic in
$J(t)$). We note  the explicit appearance of the classical boundary 
action, which is somewhat  non-trivial since 
$T_0$ only appears in  the position
of the peak of the wave packet at $t=0$. 
This can be understood as follows:  since the potential
is quadratic we know that the expectation value $\la T(t) \ra$ will 
precisely follow a classical trajectory, and in fact it is not difficult 
to show that the classical trajectory associated with the wave packet 
$\psi(T,0) \sim \exp(-a(T-T_0)^2/2$ is that of a particle 
put at $T_0$ at time $t\equ 0$ with zero momentum. Thus our choice 
of wave packet \rf{gauss} is associated with the classical path 
$T_0 \cosh \om t$. It is also worth to note that Gaussian wave packets 
at a specific time $t$ 
remain Gaussian in $T$ at any time since the propagator is Gaussian.
However, the width can change in time and the Gaussian wave packet for 
the inverted harmonic oscillator spreads so fast that even if the 
peak follows a classical trajectory, coming in from $T \equ \infty$
and being reflected at position $T_0$ at $t \equ 0$, 
finally to move back to $T\equ \infty$, it has at any time 
a  non-vanishing tail at the other side of the potential $V(T)=-T^2/2$.

Finally note that the quantum corrections have introduced 
an imaginary part in the quantum boundary action. From the 
point of view of  quantum mechanics of the (harmonic or) inverted 
harmonic oscillator it just reflects that the presence of 
a source $J(t)$ can change the wave function $\psi(T,t)$ which is
a solution corresponding to the Hamiltonian 
\rf{j2} without the presence of $J(t)$. Eq.\ \rf{e.oscinv} gives 
the amplitude for the system to remain in state $|\psi\ra$ after 
being perturbed by the external source $J$.

\subsection*{Conformal invariance}

It is well known \cite{polchinski,boundary,recknagel} that the
classical rolling tachyon boundary 
interaction is conformally invariant (in the sense that it defines a
BCFT and provides a consistent open string background). It is
therefore interesting to ask what happens when the quantum-mechanical
corrections are included.

We will now show that the boundary action (\ref{qaction}) is
conformally invariant at one loop. To this end we have to calculate
the $\beta$ function for a boundary interaction of the form:
\eq
\label{e.biloc}
\int d\tau \, T_1(t(\tau)) +\int d\tau d\tau' \, T_2(t(\tau),t(\tau'))
\eqx
At one loop the first term gives the standard contribution
$T_1-d^2/dt^2 T_1$ which obviously vanishes for the classical rolling
tachyon profile. For the second term one performs an analogous
calculation by decomposing $t(\tau)=t_0(\tau)+\xi(\tau)$, inserting
it into the action (\ref{e.biloc}) and expanding up to quadratic
order in $\xi(\tau)$. The background field $t_0(\tau)$ is chosen so
that the linear terms will vanish. The contractions are then made with 
\eq
\label{e.cor}
\cor{\xi(\tau)\xi(\tau')}=2\log |\tau-\tau'|
\eqx
regularized at coinciding points as $2\log \Lambda$. One then has to
isolate terms proportional to $\log \Lambda$. One gets a
straightforward generalization of the local terms:
\eq
\label{e.parti}
-2T_2(t,t') +\f{d^2}{dt^2} T_2(t,t')+ \f{d^2}{d{t'}^2} T_2(t,t')
\eqx
and the $\log \Lambda$ piece of
\eq
\label{e.partii}
2\f{d}{dt} \f{d}{dt'} T(t,t') \cor{\xi(\tau)\xi(\tau')}
\eqx
Let us first discuss the continous part of
(\ref{Da}). Eq. (\ref{e.parti}) then vanishes trivially, while the
integral of (\ref{e.partii}) will not give a divergence as the
logarithmic singularity in (\ref{e.cor}) is integrable. Hence it will
not generate a contribution to the $\beta$ function. 
The discontinous part 
\eq
D_a^{discont.}(t,t')=\theta(t-t') \sinh (t-t')
\eqx
satisfies
\eq
\left( \f{d^2}{dt^2} -1 \right) D_a^{discont.}(t,t')=\dl(t-t')
\eqx 
Therefore (\ref{e.parti}) gives $2\dl(t-t')$, while 
(\ref{e.partii}) gives
\eq
2\left\{ -2 \dl(t-t') \cosh(t-t') -\th(t-t') \sinh(t-t') -\dl'(t-t')
\sinh(t-t') \right\} \cor{\xi(\tau)\xi(\tau')}
\eqx
Only the first and last terms will contribute to the $\beta$ function
and will give $-2\dl(t-t')$ which cancels the contribution of
(\ref{e.parti}).

From the above we see that the quantum corrections do not spoil
conformal invariance at one loop. It would be very interesting to
verify if the result extends to higher orders in $\al'$. In fact the
properties of field theories with nonlocal (bilocal) boundary
interactions seems to be largely unexplored (for a first study see
\cite{liwitten}), yet such theories seem to appear here quite
naturally. Moreover such types of boundary interactions can have quite
nontrivial properties like mimicking closed strings \cite{us}
etc. Care must also be taken, however, as there are examples of
bilocal boundary interactions which are $SL(2,{\mathbb{C}})$ invariant
and yet break full Virasoro invariance \cite{us}.

\subsection*{The quantum decay rate}

The starting point of the calculation of the decay
of the D-brane is the disk amplitude. As 
shown in \cite{llm} the amplitude to decay into 
closed strings tachyons is (in analogy with \rf{j10} and
\rf{j11}) given by 
\beq\label{j30}
A(E) = \int dt\; e^{iEt}\, \rho(t)
\eeq
where $\rho(t)$ is obtained as (``tilde'' means no zero mode)
\beq\label{j31}
\rho(t) = \langle e^{i\int_B d \tau T_0 \cosh(t+\tilde{X}_0(\tau))}\rangle_{disk'}
\eeq
The corresponding expression for a {\it quantum} D-brane
is now obtained from \rf{qaction} by the replacement
\beq\label{j32}
e^{iT_0\int_B d \tau \; \cosh(t+\tilde{X}_0(\tau))} \to
e^{i S^{(qba)}(t+\tilde{X}_0)}
\eeq 
where the quantum boundary action $S^{(qba)}(t+\tilde{X}_0)$ is 
\beq\label{qba1}
 T_0\int_B d \tau  \cosh(t \pl \tilde{X}_0(\tau)) +\f{1}{2} 
\int_B \int_B d\tau_1 d\tau_2 \; D_a(t\pl \tilde{X}_0(\tau_1),t \pl \tilde{X}_0(\tau_2))
\eeq
We thus have a modified $\rho(t)$, much along the lines of the 
2d case:
\beq\label{qrho}
\rho_{quan}(t) = \la e^{i S^{qba}(t+\tilde{X}_0)}\ra_{disk'} =
\int {\cal D} \tilde{X}_\m  \; e^{i S^{qba}(t+\tilde{X}_0)} \;
e^{i \int_{disk} d^2 z \, \prt \tilde{X}_\m \prt {\tilde{X}}^\m},
\eeq  
and a modified emission amplitude
\beq\label{qemis}
A_{quan}(E) = \int dt \; e^{iEt}\, \rho_{quan}(t)
\eeq
{\it In principle} one can just apply the substitution \rf{j32}
and perform the functional average in the calculation 
of $\rho_{quan}(t)$. The simplest 
way to do this was outlined in \cite{finn} for the 
classical solution $T_{cl}(t) = T_0 e^t$ (the half-brane). However,
the method can also be used for $T_{cl}(t) = T_0 \cosh t$. Technically 
it is somewhat more complicated since one will encounter contraction 
of exponentials $e^{\tilde{X}_0(\tau_1)}$ and $e^{-\tilde{X}_0(\tau_2)}$, and 
these result in singularities when points 
coincide on the boundary. However, the regularization proposed 
in \cite{boundary} allows one to reproduce the result first 
derived by Sen \cite{sen}
\beq\label{j33}
\rho(t) = \f{1}{1+\sin(\pi T_0)\; e^t}  + \f{1}{1+\sin(\pi T_0)\; e^{-t}} -1.
\eeq
We mention this because one encounters no further technical obstacles when  
using the substitution \rf{j32} in the calculation \rf{qrho}.

Unfortunately we have been unable to perform the calculation 
explicitly to a level that a formula like \rf{j33} appears , 
i.e.\ expand the 
exponential $e^{iS^{(qba)}(t+\tilde{X}_0)}$, 
do all the Wick contractions of the exponentials 
{\it and}  sum the corresponding series. 
But we note the following: the qualitative form of 
$\rho(t)$ can be understood without calculations: the fast 
oscillations of $e^{i \int_B d\tau \cosh(t+\tilde{X}(\tau))}$ caused 
$\rho(t)$ to vanish for $t \to \pm \infty$. The same is likely 
to be true for $\rho_{quan}(t)$ since in addition we have
an exponentially decaying part in \rf{j32}. However, the implications
for the decay amplitude $A_{quan}(E)$ given by \rf{qemis} is less clear since 
the integral for large $E$ will be dominated by the pole closest 
to the real axis. This is the reason it is not enough to perform 
an expansion of  $\rho_{quan}(t)$ in powers of $e^t$ to finite order
(which can indeed be done, as mentioned). One needs more  precise
information in order to determine the large $E$ behavior of the 
Fourier transform $A_{quan}(E)$. 

Thus so far we cannot state that emission of 
large energy closed string states is damped by the quantum  
nature of the tachyon as was the case in 2d string theory, 
although it is likely to be true. In any case it would be very
interesting to perform such a calculation.

\subsection*{Inclusion of momentum modes}

So far we have restricted ourselves to tachyons with no momentum dependence
(the tachyon zero-mode). Let us now examine how the quantum
treatment of the momentum modes influences the picture.

As we have already seen in formula \rf{s_t}, we obtain a
set of inverted harmonic oscillators (with $k^2<1$) and
a set of normal ones (modes with $k^2>1$).
A natural wave function to consider is a Gaussian wave packet (with
$a=|\om|$) for each of the modes. For simplicity we set the initial
`position' $T_0$ to zero. This kind of wave function 
illustrates the weirdness of dealing with tachyons: 
the higher and lower modes behave completely
differently - the higher modes are  just in their stationary
vacuum state while the lower ones spread out due to the inverted
harmonic potential. 

The boundary action for an open string is given by \rf{ba} 
so  we have to calculate
\eq
S_b^{quantum}= \prod_{\vec{n}} \cor{e^{\int d\tau \f{1}{(2\pi
L)^{\f{p}{2}}} \phi^c_{\vec{n}} (t) \cdot \cos kx}}_\psi \cdot \cor{e^{\int
d\tau \f{1}{(2\pi L)^{\f{p}{2}}} \phi^s_{\vec{n}} (t) \cdot \sin kx}}_\psi 
\eqx
The result follows from \rf{Dom1} and \rf{Dom}. 
For the higher, normal (cosine) modes one obtains
\eq\label{high}
e^{\f{i}{2} \int d\tau d\tau' \f{1}{(2\pi L)^{p}} \cos kx \cos kx'
\left[\th(t-t')\f{\sin \om(t-t')}{\om} +\f{i}{2}
\f{\cos\om(t-t')}{\om}\right] } 
\eqx
where $\om=\sqrt{k^2-1}$.  
The result for the sine modes is similar. 
Note that \rf{high} contains {\it no} 
dependence on the time $t$ which enters into the calculation of 
$\rho_{quan}(t)$ in accordance with the fact that we are just using 
the standard vacuum wave functions for the higher modes. 

For the lower, tachyonic  modes we obtain from \rf{Dom}:
\eq\label{low}
e^{\f{i}{2} \int d\tau d\tau' \f{1}{(2\pi L)^{p}} \cos kx \cos kx'
\left[\th(t-t')\f{\sinh \kap(t-t')}{\kap} +\f{i}{2}
\f{\cosh\kap(t+t')}{\kap}\right] }
\eqx
with $\kap=\sqrt{1-k^2}$.

Now we have to perform a summation over $k$ in the exponential. The
volume factor and summation transforms into an integral, while the
$\om$ in the denominator gives the standard Lorentz-invariant phase
space measure. The dominant behavior for large $t,t'$ comes from 
second piece in the formula for the
lower, tachyonic modes. We thus have to evaluate:
\eq
-\f{1}{4}\int_{k^2<1} \f{d^{p}k}{\sqrt{1-k^2}}
\cos(\vec{k}\cdot(\vec{x}-\vec{x'}))\, \cosh(\sqrt{1-k^2}(t+t')) 
\eqx 
This integral can be performed analytically in the  
small  $|x-x'|$ region and we obtain
\eq
const \cdot \f{I_{\f{p-1}{2}}(t+t')}{(t+t')^{\f{D-2}{2}}}
\eqx
For large $t,t'$ this behaves exponentially like the 
zero mode (in fact it is the dominant zero mode which 
gives this contribution) -- so the qualitative
features of the result, like the exponential suppression of the
partition function/amplitudes at large times, is similar as in the
zero-mode case and we do not expect any additional surprises coming
from the space-dependent tachyonic modes.

\subsection*{Discussion}

As mentioned in the introduction the considerations in this paper
were motivated on the one hand by the desire to understand more about
the nature of quantum D-branes, and on the other hand by the very
explicit quantum effects which could be analyzed quantitatively for
D0-branes in 2d string theory \cite{aj}. Some generic aspects have
emerged from this investigation.  

The first is that the 
quantum nature $T_{cl}(t) \to \psi(T,t)$ makes it impossible 
to ignore the ``other side'' of the potential $V(T)=-T^2/2$. As discussed
above the wave packets always have a tail on the other side of the potential.
This  is bad since the ``other side'' of the {\it full} tachyonic
potential in bosonic string theory  is unbounded from below 
like $T^3$. This is much worse than $-T^2$. There is no natural 
self-adjoint Hamiltonian in such a potential and most likely
there is no cure for such a theory. It makes the role of the 
conjectured minimum on the ``right side'' of the potential
somewhat dubious.  Of course one can alway confine oneself
to the study of superstrings where the potential is bounded 
from below.

Secondly, the quantum correction to the boundary action seems to
preserve conformal invariance at least to one loop order. It would be
very interesting to study these issues in more detail.

Thirdly, there will generically be an imaginary part in the quantum
boundary action. For the simplest constant mode wave function 
\rf{gauss} with $a=1$ in {\em Minkowski} space one has a term 
\eq
e^{-\f{1}{4}\int d\tau \int d\tau' \cosh(2t+X'_0(\tau)+X'_0(\tau'))},
\eqx
This very strong suppression of ``propagation'' for large 
times seems to be a genuine quantum effect 
related to the state $\psi(T,t)$ and such behavior is characteristic for 
potentials unbounded from below like the inverse harmonic potential.

The appearance of this term has an interesting consequence with
regards to the conjecture that in the tachyon vacuum open strings
disappear and only closed strings are left. The amplitude for an open
string propagating to very large times will always be exponentially
supressed with respect to the amplitude of an open string transforming
into a closed string at some finite time, even though the latter
amplitude would be penalized by a factor of $g_s$.

Finally let us note that it would also be very interesting to explore
the role of quantum backgrounds for closed strings, not necessarily
just in the context of tachyonic physics.

\subsection*{Acknowledgment}
J.A. and R.J.  acknowledge support by the
EU network on ``Discrete Random Geometry'', grant HPRN-CT-1999-00161. 
and by ``MaPhySto'', 
the Center of Mathematical Physics 
and Stochastics, financed by the 
National Danish Research Foundation. RJ was supported in part by KBN
grants 2P03B09622 (2002-2004), 2P03B08225 (2003-2006) and 1P03B02427
(2004-2007).

\end{document}